\documentclass[10pt,letterpaper]{article}
\usepackage[top=0.85in,left=2.75in,footskip=0.75in]{geometry}

\usepackage{amsmath,amssymb}

\usepackage{changepage}

\usepackage[utf8x]{inputenc}

\usepackage{textcomp,marvosym}

\usepackage{cite}

\usepackage{nameref,hyperref}

\usepackage[right]{lineno}

\usepackage{microtype}
\DisableLigatures[f]{encoding = *, family = * }

\usepackage[table]{xcolor}

\usepackage{array}

\newcolumntype{+}{!{\vrule width 2pt}}

\newlength\savedwidth

\raggedright
\usepackage[aboveskip=1pt,labelfont=bf,labelsep=period,justification=raggedright,singlelinecheck=off]{caption}

\makeatletter
\renewcommand{\@biblabel}[1]{\quad#1.}
\makeatother

\usepackage{lastpage,fancyhdr,graphicx}
\usepackage{epstopdf}
\fancyheadoffset[L]{2.25in}
\fancyfootoffset[L]{2.25in}
\begin{document}
\vspace*{0.2in}

\begin{flushleft}
{\Large
\textbf\newline{Influence of agent's self-disclosure on human empathy} 
}
\newline
\\
Takahiro Tsumura\textsuperscript{1,2*},
Seiji Yamada\textsuperscript{2,1}
\\
\bigskip
\textbf{1} Department of Informatics, The Graduate University for Advanced Studies, SOKENDAI, Tokyo, Japan
\\
\textbf{2} National Institute of Informatics, Tokyo, Japan
\\
\bigskip

%
%





* takahiro-gs@nii.ac.jp

\end{flushleft}
\section*{Abstract}
As AI technologies progress, social acceptance of AI agents, including intelligent virtual agents and robots, is becoming even more important for more applications of AI in human society. 
One way to improve the relationship between humans and anthropomorphic agents is to have humans empathize with the agents. 
By empathizing, humans act positively and kindly toward agents, which makes it easier for them to accept the agents. 
In this study, we focus on self-disclosure from agents to humans in order to increase empathy felt by humans toward anthropomorphic agents. 
We experimentally investigate the possibility that self-disclosure from an agent facilitates human empathy. 
We formulate hypotheses and experimentally analyze and discuss the conditions in which humans have more empathy toward agents. 
Experiments were conducted with a three-way mixed plan, and the factors were the agents' appearance (human, robot), self-disclosure (high-relevance self-disclosure, low-relevance self-disclosure, no self-disclosure), and empathy before/after a video stimulus. 
An analysis of variance (ANOVA) was performed using data from 918 participants. We found that the appearance factor did not have a main effect, and self-disclosure that was highly relevant to the scenario used facilitated more human empathy with a statistically significant difference. 
We also found that no self-disclosure suppressed empathy. 
These results support our hypotheses. 
This study reveals that self-disclosure represents an important characteristic of anthropomorphic agents which helps humans to accept them.



\section*{Introduction}
Humans live in society and use various tools, and artifacts are sometimes treated as if they were human beings. 
It is known that humans tend to treat artificial objects like humans in media equations\cite{Reeves96}. 
However, these objects are very pervasive, and humans may not consider some of them to be like humans. 
In fact, AI's problems today are related to the reliability and to the ethical usage of AI technologies. 
Ryan\cite{Ryan20} focused on trust and discussed AI ethics and the issue of people anthropomorphizing AI. 
It was determined that even complex machines like those that use AI should not be viewed as trustworthy. 
Instead, he suggested that we should ensure that organizations that use AI and the individuals within those organizations are trustworthy. 
AI ethics was also discussed in depth from an applied-ethics perspective in a study by  Hallamaa and Kalliokoski\cite{Hallamaa22}.
\\ \indent
At the same time as trust, we often empathize with artificial objects. 
Artificial objects that we empathize with include cleaning robots, pet-type robots, and anthropomorphic agents that provide services in online shopping and at help desks. 
These are already used in society and coexist with humans. 
In addition, the appearance of these agents varies depending on the application and usage environment. 
However, some humans cannot accept these kinds of agents\cite{Nomura06,Nomura08,Nomura16}. 
As agents continue to permeate society, they should have elements that humans find acceptable.
\\ \indent
In this study, we use a target empathy agent to facilitate human empathy. 
We previously focused on ``(human) empathy" as an attribute that an agent should have in order to be accepted by human society, and we aimed to create an empathy agent. 
In this study, we focused on how humans improve their relationships with agents. 
One method is to have humans empathize with the agents. 
By empathizing, humans act positively toward agents and are more likely to accept them. 
Various studies have been done on linguistic information\cite{Konrath15,Shaffer19,Tahara19}, nonverbal information\cite{Tisseron15,Yoshioka15,Okanda19}, situations\cite{O'Connell13,Zhi18,Richards18}, and relationships\cite{Stephan15,Hosseinpanah18,Giannopulu20} as factors that cause empathy.
\\ \indent
In previous studies, self-disclosure is also regarded as important among humans. 
Therefore, we thought that self-disclosure would be necessary for anthropomorphic agents to establish relationships with humans. 
We focused on self-disclosure and experimentally examined what characteristics of self-disclosure affect empathy. 
Moreover, empathy has been studied in the fields of HAI and HRI. 
However, different appearances have been used in each study. 
Hence, we decided to set appearance as a factor to compare the effect that it has on empathy. 
For this, human-like and robot-like appearances were prepared and tested as symbols of the HAI and HRI fields.
\\ \indent
In this study, we assume that empathy agents influence human empathy. 
However, we investigate only human empathy toward empathy agents, not the human capacity for empathy, since our focus is on investigating the factors that make agents acceptable to humans. 
Below, empathy in this study refers to this kind of empathy.
\\ \indent
Considering its relevance in previous studies on self-disclosure and empathy, we select self-disclosure as a factor to investigate human empathy toward agents. 
We focus on self-disclosure from agents to humans, and we conduct experiments to investigate the relationship between human empathy and agent self-disclosure, as well as the characteristics of self-disclosure that are effective in promoting empathy. 
At the same time, we investigate the relationship between anthropomorphic agents with different appearances and self-disclosure. 
\\ \indent
In the remaining sections, we propose our empathy agents for facilitating human empathy. 
Then, we cover our experiments and the results. 
Finally, we discuss the results and describe our future work. 

\subsection*{Definition of empathy}
We consider empathy to be a significant element in being accepted by humans as a member of society. 
For humans to get along with each other, it is important that they empathize with the other\cite{Gaesser13,Klimecki16}. 
\\ \indent
Empathy and the effects that it has on others have been a focus of research in the field of psychology. 
Omdahl\cite{Omdahl95} roughly classifies empathy into three types: (1) affective empathy, which is an emotional response to the emotional state of others, (2) cognitive understanding of the emotional state of others, which is defined as cognitive empathy, and (3) empathy including the above two. 
Preston and de Waal\cite{Preston02} suggested that at the heart of the empathic response was a mechanism that allowed the observer to access the subjective emotional state of the target. 
They defined the perception-action model (PAM) and unified the different perspectives in empathy. 
They defined empathy as three types: (a) sharing or being influenced by the emotional state of others, (b) assessing the reasons for the emotional state, and (c) having the ability to identify and incorporate other perspectives. 
Olderbak et al.\cite{Olderbak14} described theoretical and empirical support for the emotional specificity of empathy and presented an emotion-specific empathy questionnaire that assesses affective and cognitive empathy for six basic emotions.
\\ \indent
Although we focus on the positive effects of empathy to improve society's acceptance of AI agents in this study, empathy has been discussed from various aspects including negative effects in psychological literature. 
Bloom tried to introduce a neutral aspect of empathy by introducing not only positive influences but also negative ones\cite{Bloom16}. 
He claimed that it is possible for empathy to act as a moral guide that leads humans to irrational decision-making and relationships to violence and anger. 
Also, he claimed that we can overcome this problem by using conscious, deliberative reasoning and altruistic approaches. 
\\ \indent
Various questionnaires are used as a measure of empathy, but we examined two famous ones. 
The Ten Item Personality Inventory (TIPI) is used to investigate human personality \cite{Gosling03}. 
For the experiment of this study, since empathy may be biased by human personality, TIPI could be used as a questionnaire survey. 
The Interpersonal Reactivity Index (IRI), also used in the field of psychology, is used to investigate the characteristics of empathy\cite{Davis80}. 
Baron-Cohen and Wheelwright\cite{Baron-Cohen04} reported a new self-report questionnaire, the Empathy Quotient (EQ), for use with adults of normal intelligence. 
Lawrence et al.\cite{Lawrence04} investigated the reliability and validity of the EQ and determined its factor structure. 
Experimental results showed a moderate association between the EQ subscale and the IRI subscale. 
We ultimately decided to use only the IRI as a questionnaire for its suitability with our experiment.

\subsection*{Empathy in engineering}
Empathy has been studied not only in the field of psychology but also in the field of engineering. 
For example, empathy has received a lot of attention in the field of virtual reality. 
Bertrand et al.\cite{Bertrand18} proposed a theoretical analysis of various mechanisms of empathy practice to define a possible framework for the design of empathy training in virtual reality. 
Herrera et al.\cite{Herrera18} compared the short- and long-term effects of traditional and VR viewpoint acquisition tasks. 
They also conducted experiments investigating the role of technological immersion with respect to different types of intermediaries. 
Curran et al.\cite{Curran19} investigated empathy by showing a video from the visual perspective of a person watching a virtual reality movie. 
\\ \indent
Empathy has also been studied in the online environment. 
Pfeil and Zaphiris\cite{Pfeil07} performed a qualitative content analysis on 400 messages from a bulletin board on depression to investigate how empathy was expressed and facilitated in online communication.
Empathy is also attracting attention in product design, and Bennett and Rosner\cite{Bennett19} discussed and investigated a human-centered design process (promise of empathy) in which designers try to understand the target user in order to inform technology development.
\\ \indent
Chella et al.\cite{Chella20} discussed self-awareness and inner speech in humans and AI agents and provided an initial proposal for a cognitive architecture for implementing inner speech in robots. 
While the foundations of internal speech had been investigated primarily in the fields of psychology and philosophy, research in robotics had not yet addressed self-aware behavior. 
Therefore, after discussing self-awareness and inner speech in humans and AI agents, they proposed the above cognitive architecture. 
Their approach had an advantage in that a robot's inner speech could be heard by an external observer, and introspective and self-regulated speech could be detected.

\subsection*{Empathy in human-robot interaction}
In other fields studying empathy, humans empathize with artificial objects. 
In the fields of human-robot interaction (HRI) and human-agent interaction (HAI), empathy between humans and agents and robots is studied. The following research has been conducted in the field of HRI.
\\ \indent
Beck et al.\cite{Beck10} studied the effects of changing a robot's head position on the interpretation of emotional key poses, valence, arousal, and stances. 
The results supported the idea that body language is an appropriate medium for robots to express emotions. 
On the basis of the concept of cognitive developmental robotics, Asada\cite{Asada15} proposed ``affective developmental robotics," which produces more truly artificial empathy. 
Artificial empathy refers to AI systems (such as companion robots and virtual agents) that can detect emotions and respond empathically. 
The design of artificial empathy is one of the most essential issues in social robotics, and empathic interaction with the public is necessary to introduce robots into society.
\\ \indent
Dumouchel et al.\cite{Dumouchel17} also summarized artificial empathy. 
The relationship between humans and robots appearing in daily life was discussed. 
They suggested that the human-robot dynamics in emotional relationships need to be considered. 
Mollahosseini et al.\cite{Mollahosseini18} applied a deep neural network-based system for automatically recognizing facial expressions to the speech dialogue of social robots. 
The function was extended and enhanced beyond voice dialogue to integrate the user's emotional state into the robot's reaction.
\\ \indent
Several studies on inner speech have been conducted in the HRI field. 
Pipitone and Chella\cite{Pipitone21} investigated the potential of considering the inner speech of robots that cooperate with human partners. 
A domestic situation requiring several functional and moral requirements was simulated in a simple cooperative task. 
Their study was a novel endeavor, as only a few papers have analyzed the role of inner speech in robots, and most of them were theoretical in nature.
Geraci et al.\cite{Geraci21-2} investigated whether a robot's internal conversation affects human trust and anthropomorphism when humans and robots collaborate together. 
The results suggest that a robot's speech affects human trust. 
The results also indicated that participants' perceptions of trust and anthropomorphism toward the robot improved after interacting with the robot in the experiment.

\subsection*{Empathy in human-agent interaction}
In addition, the following research has been done in the field of HAI. 
McQuiggan et al.\cite{McQuiggan08} proposed a unified inductive framework for modeling parallel and reactive empathy, which empathy models by choosing appropriate parallel or reactive empathy expressions. 
The framework was used to facilitate empathic behavior suitable for run-time situations. 
Leite et al.\cite{Leite14} conducted a long-term survey in an elementary school to present and evaluate an empathy model for social robots that is aimed for interactions with children that occur over a long period of time. 
They measured children's perceptions of social presence, engagement, and social support. 
\\ \indent
Chen and Wang\cite{Chen19} hypothesized that empathy and anti-empathy were closely related to a creature's inertial impression of coexistence and competition within a group, and they established a unified model of empathy and anti-empathy. 
They also presented the Adaptive Empathetic Learner (AEL), an agent training method that enables evaluation and learning procedures for emotional utilities in a multi-agent system. 
Perugia et al.\cite{Perugia20} investigated which personality and empathy traits were related to facial mimicry between humans and artificial agents. 
They focused on the humanness and embodiment of agents and the influence that they have on human facial mimicry. 
As a result, mimicry was found to be affected by the embodiment that an agent has, but not by its humanness. 
It was also correlated with both individual traits indicating sociability and empathy and with traits favoring emotion recognition.
\\ \indent
Paiva defined the relationship between humans and empathetic agents, called empathy agents in HAI and HRI studies. 
As a definition of empathy between agents/robots and humans, Paiva represents empathy agents in two different ways: targeting empathy and empathizing with observers\cite{Paiva04,Paiva11,Paiva17}.

\subsection*{Self-Disclosure in Psychology}
Self-disclosure has also been a focus of research in the field of psychology. 
Jourard\cite{Jourard71} presented the Jourard Self-Disclosure Questionnaire (JSDQ), a self-disclosure classification and questionnaire. 
Attitudes, opinions, interests, study and work, personality, economy, and body were listed as categories. 
Carpenter and Freese\cite{Carpenter79} measured self-presentation intimacy and internality, Derlega and Berg\cite{Berg87} focused on the association between responsiveness and self-disclosure, and Laurenceau JP\cite{Laurencea98} suggested that both self-disclosure and partner responsiveness contribute to the experience of intimacy in interactions. 
\\ \indent
One study related to self-disclosure is the study of inner speech. 
Morin\cite{Morin05} reviewed past and current literature on the link between self-awareness and inner speech. 
Among multidimensional views of self-knowledge, he showed that inner speech accounts for half of the linkages between various elements and plays a fundamental role.
In addition, Morin\cite{Morin20} further studied internal speech. 
He considered inner speech as creating psychological distance between the self and the mental events experienced by the self, that the self represents a problem, that self-information functions as a problem-solving device for resolution, and that it is possible to label internal aspects of the self that are otherwise difficult to recognize objectively. 
We emphasize that inner speech and imagined interactions (IIs) are not identical and differ in important ways. 
Therefore, although IIs and inner speech intersect, their overlap is quite limited, so it is possible to investigate one over the other.
\\ \indent
Lockwood et al.\cite{Lockwood17} used self-reported measurements of empathy and apathy motivation in a large sample of healthy people to test whether more empathic people were more motivated. 
The actual self-disclosure reflected in interpersonal relationships has been investigated in a few studies. 
Therefore, Kreiner and Levi-Belz\cite{Kreiner19} designed new objective and dynamic measurements to evaluate self-disclosure and stable self-disclosure characteristics. 
Oh Kruzic et al.\cite{OhKruzic20} focused on how the face and upper-body nonverbal channels contribute individually and collaboratively via avatars in virtual environments. 
Lee et al.\cite{Lee20} found that including self-disclosure from chatbots when they interacted with humans had an effect on improving participants' perceptions of intimacy and enjoyment. 
Pan et al.\cite{Pan20} examined the effect of exposure to online support-seeking posts containing different levels of self-disclosure depth (baseline, peripheral, core) affecting the quality (person-centeredness and politeness) of participants' messages providing support.

\section*{Materials and methods}
\subsection*{Ethics Statement}
The protocol was approved by the ethics committee of the National Institute of Informatics (13, April, 2020, No. 1). 
All studies were carried out in accordance with the recommendations of the Ethical Guidelines for Medical and Health Research Involving Human Subjects provided by the Ministry of Education, Culture, Sports, Science and Technology and Ministry of Health, Labour and Welfare in Japan. Written informed consent was provided by choosing one option on an online form: ``I am indicating that I have read the information in the instructions for participating in this research. 
I consent to participate in this research." All participants gave informed consent. 
After that, they were debriefed about the experimental procedures. 

\subsection*{Hypotheses}
The purpose of this study was to investigate whether it is possible to elicit more human empathy when an empathy agent performs self-disclosure related to a particular situation in an interaction with a human.
In this experiment, three types of self-disclosure topics were prepared, work, hobby, and weather or land, in order of relevance to the situation during a conversation about work. 
In addition, the appearances of the agents were human-like and robot.
This objective is an important condition for humans and agents to cooperate in society.
If our hypothesis is supported, this research can help develop agents that are more acceptable to humans.
\\ \indent
Based on the above, we considered two hypotheses. 
Experiments were conducted to investigate these hypotheses.

\begin{enumerate}
    \item[H1:] Of the three types of self-disclosure from empathy agents (high-relevance self-disclosure, low-relevance self-disclosure, no self-disclosure), high-relevance self-disclosure can facilitate empathy the best of them, and no self-disclosure suppresses empathy.
    \item[H2:] In interacting with agents, appearance factors have little impact on promoting empathy through self-disclosure.
\end{enumerate}

\subsection*{Experimental procedure}
\begin{figure}[tbp]
    \hspace{-1cm}
    \includegraphics[scale=0.45]{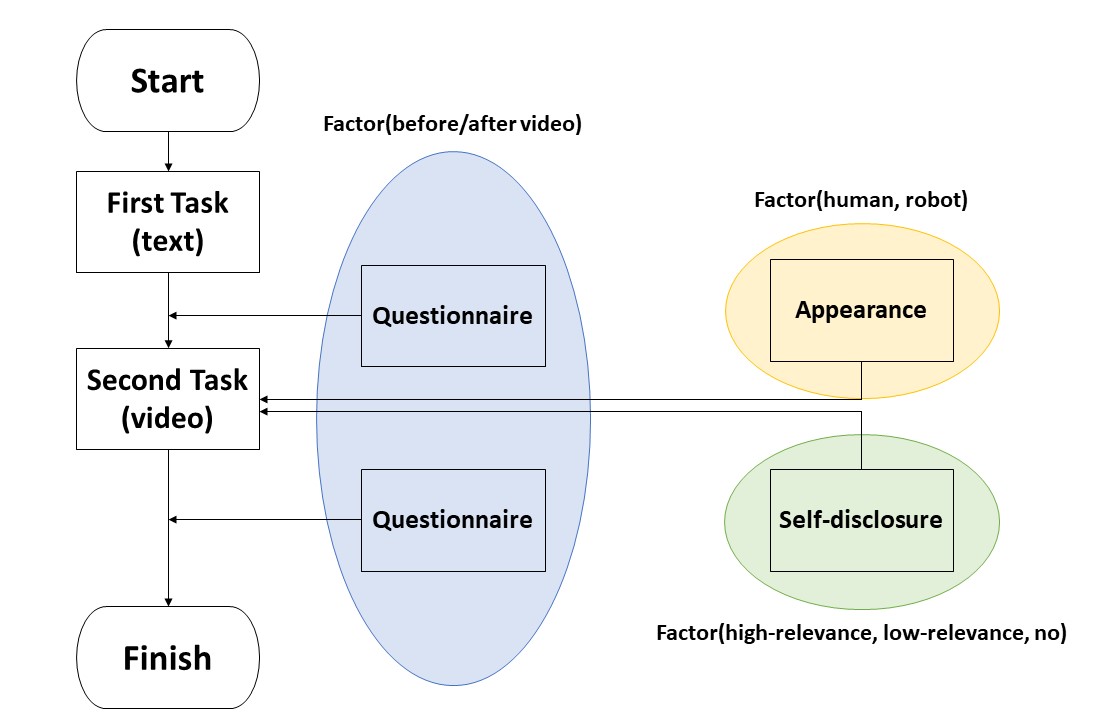}
    \vspace{1mm}
    \caption{Flowchart of the experiment.}
    \label{fig1}
\end{figure}

The experiments were conducted in an online environment. 
The online environment used in this experiment has already been used as one experimental method\cite{Davis99,Crump13,Okamura20}. 
The flowchart of this experiment is shown in Figure~\ref{fig1}.
Participants performed two tasks. 
Below, we describe the two tasks.
\\ \indent
In the first task, the participants are asked to read a simple abstract prepared in advance in text format so that they could understand the relationship with the agent. 
They were only to read while imagining the agent. 
After they read the abstract, the empathy that they felt for the agent was tabulated in a questionnaire survey. 
In this task, participants did not judge the appearance of the agent or self-disclosure.
\\ \indent
In the second task, a three minute video that was made from the content of the first task was shown to the participants. 
The agent in the video spoke silently to the participants through a text box. 
The reason for the silence is that sound may affect the facilitation of empathy. 
In addition, gestures were performed at the same timing under all conditions. 
Participants interacted with the agent under any one of a total of six conditions that combined two factors: appearance (human, robot) and self-disclosure (high-relevance self-disclosure, low-relevance self-disclosure, no self-disclosure). 
The control state was a condition of no self-disclosure. 
The content except for the case of self-disclosure was the same, so it was possible to investigate the promotion of empathy due to the difference in self-disclosure. 
After that, as with the first task, the empathy felt toward the agent was tabulated in a questionnaire survey. 
After completing all the tasks, we asked them to write their impressions of the experiment in a free description. 
\\ \indent
Thus, the independent variables were self-disclosure (high-relevance, low-relevance and no self-disclosure), agent's appearance (human, robot), and before/after stimulation (before, after video). 
The dependent variables were human empathy and human empathic response. 
\\ \indent
The experiments were conducted with a three-factor mixed-plan. 
The number of between-participant factors was two, appearance and self-disclosure, and the within-participant factor was the empathy values before/after video stimulus used to measure the change in empathy. 
The number of levels of each factor was two for appearance (human, robot), three for self-disclosure (high-relevance self-disclosure, low-relevance self-disclosure, no self-disclosure), and two for stimulation (before, after). 
Although there were 12 levels in total, participants were asked to join in only 1 of 6 different experiments due to the within-participant factor.

\subsection*{Participants}
We used Yahoo! Crowdsourcing to recruit participants, and we paid 70 yen (= 0.67 dollars US) to each participant as a reward. 
We created web pages for the experiments by using Google Forms, and we uploaded the video created for the experiment to YouTube and embedded it.
\\ \indent
All participants had an understanding of Japanese. 
There were a 1011 participants in total. However, since there were 32 participants who gave inappropriate answers, their data was excluded as erroneous, so the final total was 979. 
To judge whether answers were inappropriate in the experiments, we judged answers as inappropriate when the changes in the empathy values before/after video were the same for all items or when only one item changed\cite{Schonlau15,Leiner19}. 
After that, as a result of using Cronbach's $\alpha$ coefficient for the reliability of the questionnaire, the coefficient was determined to be 0.7155 to 0.8201 under all conditions. 
\\ \indent
For the analysis, 153 people were analyzed under each of six conditions in the order of participation. 
Therefore, the total number of participants used in the analysis was 918. 
The average age was 45.51 years (S.D. = 11.25), with a minimum of 15 years and a maximum of 86 years. 
In addition, there were 505 males and 413 females.

\subsection*{Questionnaire}
In this study, we used a questionnaire related to empathy that has been used in previous psychological studies.
To investigate the characteristics of empathy, we modified the Interpersonal Reactivity Index (IRI) to be an index for anthropomorphic agents. 
The main modifications were changing the target name and changing the question text to the past tense. 
In addition, the number of items on the IRI questionnaire was modified to 12; for this, items that were not appropriate for the experiment were deleted, and similar items were integrated. 
The same questionnaire was used for both tasks. 
Since both of the questionnaires used were based on IRI, a survey was conducted using a 5-point Likert scale (1: not applicable, 5: applicable). 
\\ \indent
The questionnaire used is shown in Table~\ref{table1}. 
Since Q4, Q9, and Q10 were reversal items, the points were reversed during analysis. 
Q1 to Q6 were related to affective empathy, and Q7 to Q12 were related to cognitive empathy.
Only the second task had one additional question, which is shown in Table~\ref{table1} as BeQ. 
This was an item for investigating the empathic response of the participants, and they answered this question with either yes or no.
Participants answered a questionnaire after completing the first task and the second task. 

\renewcommand{\arraystretch}{1.5}
\begin{table}[tbp] 
    \caption{Summary of questionnaire used in this experiment}
    \vspace{1mm}
    \begin{adjustwidth}{-2.0in}{0in} 
    \centering
    \scalebox{1.0}{
    \begin{tabular}{l}\hline 
        \textbf{Affective empathy}\\ \hline
        \textbf{Personal distress}\\
        Q1: The agent experienced an emergency, and you became anxious and uncomfortable.\\
        Q2: You didn't know what to do when the agent was emotional.\\
        Q3: When you saw someone in need of immediate help, you were confused and didn't know what to do.\\
        \textbf{Empathic concern}\\
        Q4: You didn't feel sorry to see the agent in trouble.\\
        Q5: Seeing that the agent was being used in a good way by others made you want to protect that agent.\\
        Q6: You were deeply moved by the story of the agent and what happened.\\\hline
        \textbf{Cognitive empathy}\\ \hline
        \textbf{Perspective taking}\\
        Q7: You tried to look at both the agent position and the human position.\\
        Q8: You tried to get to know the agent well and imagined how things were seen from the agent.\\
        Q9: When you thought you were right, you didn't listen to the agent.\\
        \textbf{Fantasy scale}\\
        Q10: You were objective, not drawn into the agent's story or what happened.\\
        Q11: You imagined what it would be like if something that happened to the agent happened to you.\\
        Q12: You got deep into the feelings of the agent.\\\hline
        \textbf{Empathic response}\\ \hline
        BeQ: Finally, the agent has asked you to lend it some money. What would you do?\\ \hline
    \end{tabular}}
    \label{table1}
    \end{adjustwidth}
\end{table}
\renewcommand{\arraystretch}{1.0}

\subsection*{Agents' appearance}
\begin{figure}[tbp]
    \includegraphics[scale=0.3]{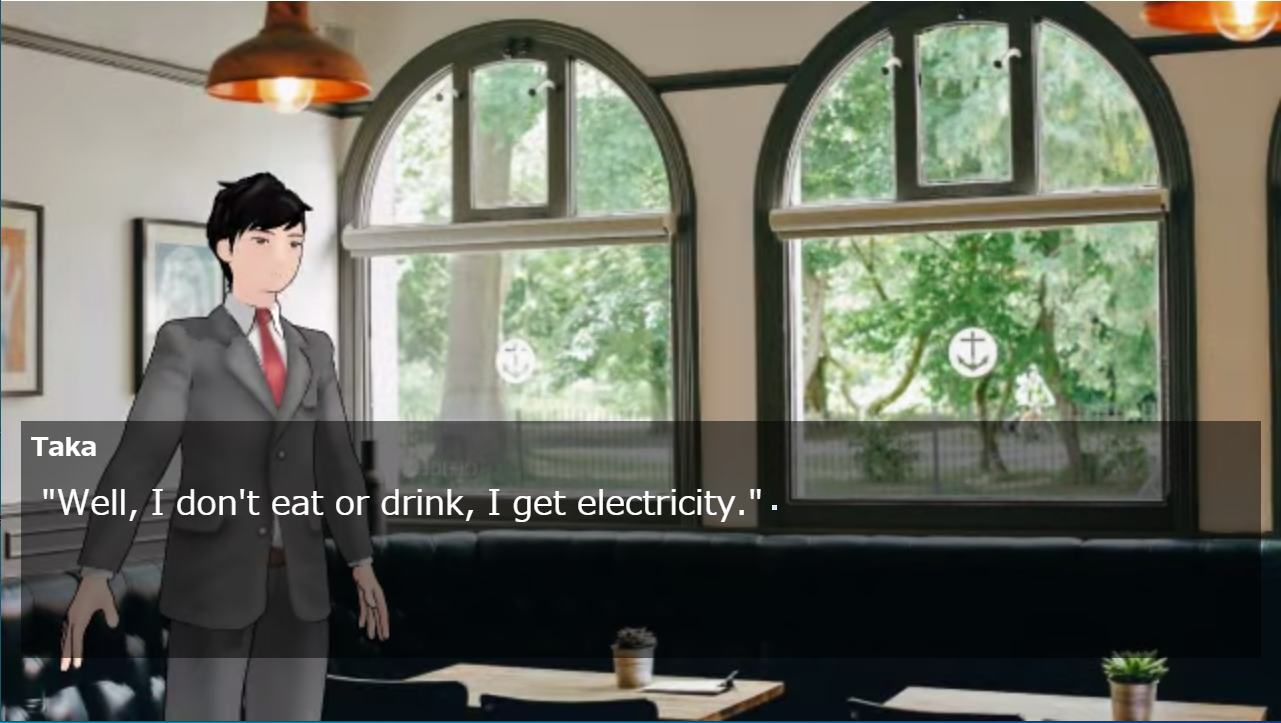}
    \vspace{1mm}
    \caption{Scene of video when appearance was human}
    {Part where human-like agent and participants interacted.}
    \label{fig2}
\end{figure}

\begin{figure}[tbp]
    \includegraphics[scale=0.3]{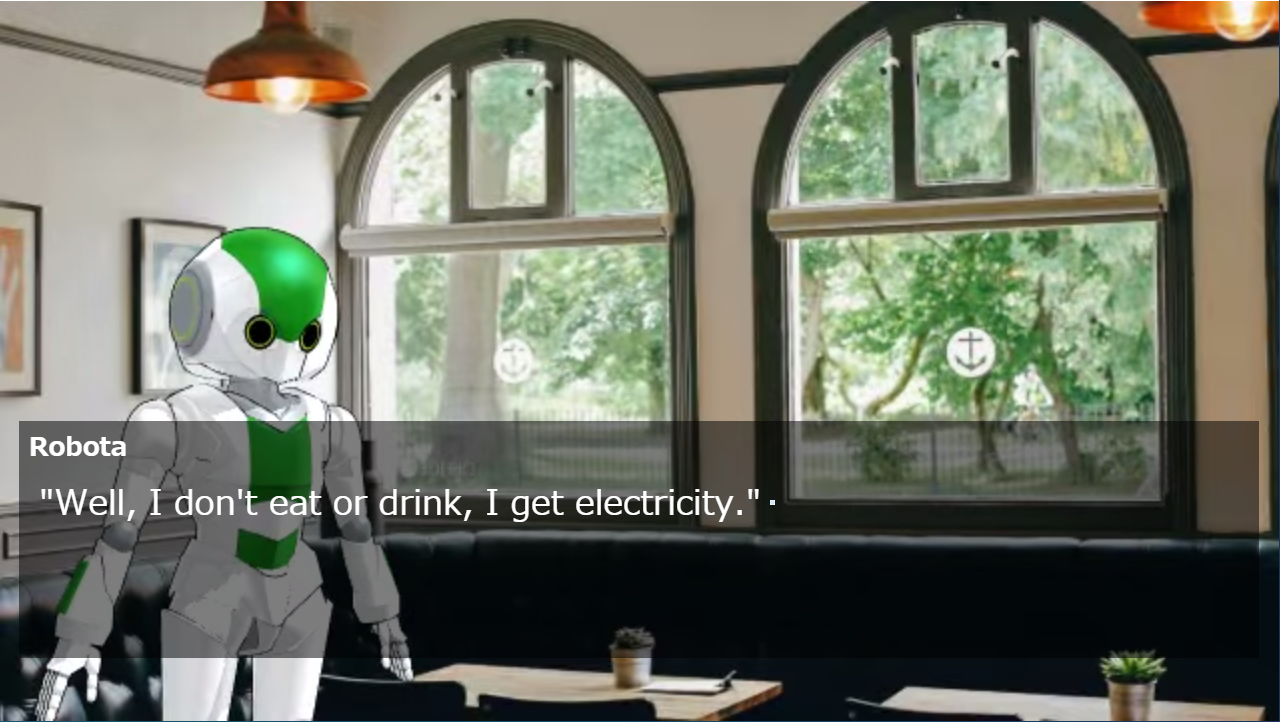}
    \vspace{1mm}
    \caption{Scene of video when appearance was human}
    {Part where robot agent and participants interacted.}
    \label{fig3}
\end{figure}

In this experiment, two types of agent appearances were prepared. 
These agents were run on MikuMikuDance (MMD)\footnote{https://sites.google.com/view/evpvp/}.
MMD is a software program that runs 3D characters.
\\ \indent
Figure~\ref{fig2} and Figure~\ref{fig3} show robot-like and human-like appearances. 
The purpose of preparing two appearances was to investigate one of our hypotheses, that is, that appearance factors do not affect the promotion of empathy through self-disclosure. 
Agent gestures included tilting the left and right arms and neck, and both agents operated at the same timing in the scenario. 
As for facial expressions, the human slightly moved their eyes and mouth, but the robot moved only their eyes.

\subsection*{Agent's self-disclosure}
\begin{figure}[tbp]
    \hspace{-1cm}
    \includegraphics[scale=0.3]{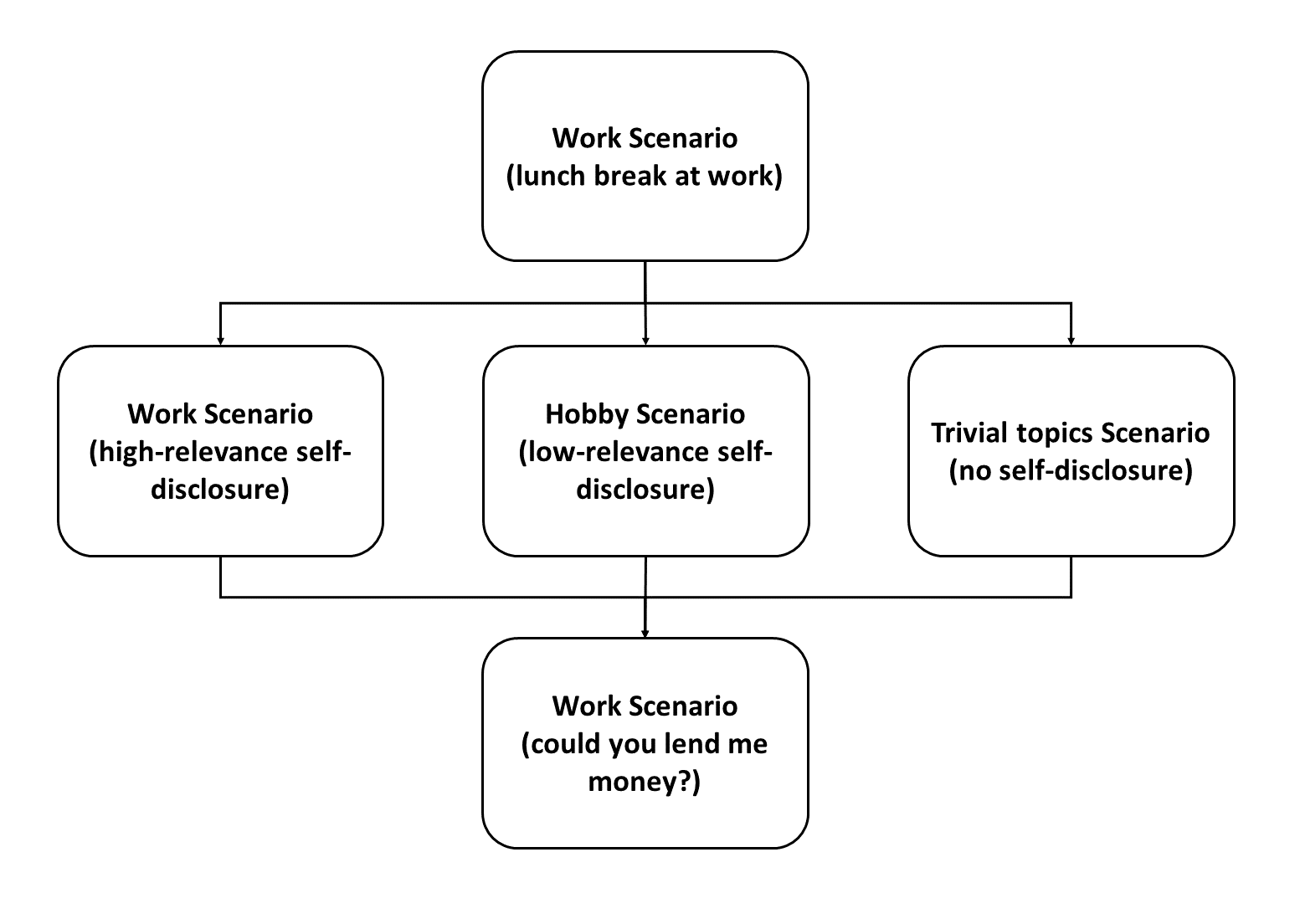}
    \caption{Flowchart of scenario.}
    \label{fig4}
\end{figure}

The scenario was that the participants were chatting at a cafe during a lunch break as a colleague at the agent's workplace. 
All scenarios started with a common content.
After that, there was a content that included self-disclosure of the agents in each condition.
Finally, the scenario ended with the common content.
A flowchart of the scenario is shown in Figure~\ref{fig4}.
\\ \indent
The common scenario involved a conversation about the nature of the job. 
Self-disclosure in this experiment referred to personal information (e.g., work, hobby) about the agent. 
In this experiment, self-disclosure was classified into the above three types, and as shown in the figure, self-disclosure was defined in accordance with its relevance to the common scenario. 
Therefore, self-disclosure in this study was most relevant for content about work. 
Stories about hobbies were less relevant because they involved self-disclosure not related to the common scenario. 
Finally, to unify the participants' interaction time, no self-disclosure was defined as talking about the weather or the land.
\\ \indent
The agents' self-disclosure was classified into three types: high-relevance, low-relevance, and no self-disclosure. 
At this time, all the content spoken by the agent was set to be neutral by sentiment analysis. 
The analysis was performed for all of the scenario in Python, and the numbers ranged from 0.075 to 0.190. 
Since this analysis ranges from -1 to 1, 0.075 to 0.190 can be classified as neutral.
\\ \indent
The difference of the scenario was in the content of the self-disclosure of the agent. 
Since the agent spoke about his own work in the cafe scenario, the content in the case of high-relevance self-disclosure was related to work. 
The low-relevance self-disclosure was related to the hobby of the agent. 
To adjust the video time for the case of no self-disclosure, the agent spoke about trivial topics, such as the weather and local area information, to consume time. 
All of the videos were about 3 minutes.
All scenarios are described in the Appendix.
A manipulation check was performed to ensure that the self-disclosure used in this study was as expected. 

\subsection*{Manipulation check: Relevance of self-disclosure and degree of self-disclosure}
We created two versions of the scenario: a common scenario and ones in which each type of self-disclosure were performed under this common scenario. 
Please review the scenario in the appendix.
It was necessary to check that the types we prepared were what we had intended. 
By performing a manipulation check, we confirmed that the three types (high-relevance self-disclosure, low-relevance self-disclosure, no self-disclosure) gave the intended impression in the cafe scenario we used. 
\\ \indent
As a manipulation check, we conducted an experiment to investigate the relationship between the scenarios and self-disclosure and the degree of self-disclosure. 
We asked the participants to read only the text of the common scenario (scenario 1) and then read the scenario for each self-disclosure condition (scenario 2). 
Afterward, they answered a questionnaire. 
\\ \indent
There were two questions (relevance of self-disclosure: Were the two scenarios related to each other?; degree of self-disclosure: How much self-disclosure was in scenario 2?). 
A 7-point Likert scale was used (1: unrelated, no self-disclosure, 7: related, high self-disclosure). 
This was a one-factor experiment between participants, and there were three levels of self-disclosure. 
The analysis was a one-way ANOVA among one-factor participants. 

\subsubsection*{Manipulation check: Participants}
We used Yahoo! Crowdsourcing JAPAN to recruit participants, and we paid 32 yen (= 0.30 dollars US) to each participant as a reward. 
We created web pages for the experiment by using Google Forms, and we uploaded the videos created for the experiment to YouTube and embedded them. 
All participants had an understanding of Japanese. 
There was a total of 154 participants. 
The average age was 44.16 years (S.D. = 9.559), with a minimum of 20 years and a maximum of 63 years. 
In addition, there were 115 males and 39 females.

\subsubsection*{Manipulation check: Result of analysis}
For multiple comparisons, we used Holm's multiple comparison test to examine whether the results were significant. 
Since the factors were significant in the results of each questionnaire, the main effect was investigated [relevance of self-disclosure: $F$(2,151) = 76.70, $p$ = 0.0000, $\eta^2_p$ = 0.5040, degree of self-disclosure: $F$(2,151) = 102.44, $p$ = 0.0000, $\eta^2_p$ = 0.5757]. 
The results of the analysis indicated that the main effect was significant, so the results of the multiple comparisons were investigated.
\\ \indent
The high-relevance self-disclosure conditions were found to be highly relevant to the most common scenario. 
In addition, relevance of self-disclosure showed a significant difference in the combination of all three levels, they were considered to be related to the common scenario in the order of high-relevance self-disclosure (mean = 5.920, S.D. = 1.426) \textgreater low-relevance self-disclosure (mean = 3.510, S.D. = 1.870) \textgreater no self-disclosure (mean = 2.132, S.D.=1.359). 
From this, it was found that the cafe scenario designed by us had a degree of relevance to self-disclosure.
\\ \indent
Next, it was observed that the degree of self-disclosure under each self-disclosure condition seemed to be in the order of high-relevance self-disclosure (mean = 5.820, S.D. = 0.9624) \textgreater low-relevance self-disclosure (mean = 5.010, S.D. = 0.9693) \textgreater no self-disclosure (mean = 2.585, S.D. = 1.550). 
In this experiment, the maximum evaluation that could be given was 7 points, so the average for high- and low-relevance self-disclosure was 4 points or more, and that for the case of no self-disclosure was less than 4 points. 
From the above results, it was judged that the content of the high- and low-relevance disclosure was self-disclosure.
\\ \indent
Also, as a result of a post-hoc analysis, the effect size of the relevance of self-disclosure was 1.008, and the effect size of the degree of self-disclosure was 1.165. 
The power of the relevance of self-disclosure was 1.000, and the power of the degree of self-disclosure was 1.000. 
It was also found that both the degree of relevance to the scenario and the degree of self-disclosure were effective. 
This manipulation check was able to objectively confirm the relevance of self-disclosure and the degree of self-disclosure in the cafe scenario we created. 
Our study was conducted using this scenario.

\subsection*{Analysis method}
We employed an ANOVA for a three-factor mixed-plan. 
ANOVA has been used frequently in previous studies and is an appropriate method of analysis with respect to the present study. 
The between-participant factors were two levels of appearance and three levels of self-disclosure.
There were two levels for the within-participant factor, before/after video.
\\ \indent
From the results of the participants' questionnaires, we investigated how self-disclosure and appearance affected the promotion of empathy as factors that elicit human empathy. 
The values of empathy aggregated in the first task and the second task were used as the dependent variable. 
For the empathic response, the Yes/No answer was replaced with a 1/0 dummy variable, and an ANOVA between two factors was then performed. 
R (ver. 4.1.0), statistical software, was used for the ANOVA and multiple comparisons in all analyses in this paper.

\section*{Results}
\renewcommand{\arraystretch}{1.1}
\begin{table}[tbp]
\caption{Analysis results of ANOVA}
\vspace{1mm}
\begin{adjustwidth}{-1.0in}{0in} 
\scalebox{1.0}{
\begin{tabular}{cllll}\hline
& \multicolumn{1}{c}{Factor} & \multicolumn{1}{c}{\em{F}} & \multicolumn{1}{c}{\em{p}} & \multicolumn{1}{c}{$\eta^2_p$}\\ \hline
& Appearance & 1.608 & 0.2051 \em{ns} & 0.0018 \\ 
& Self-disclosure & 9.616 & 0.0001 *** & 0.0207\\
Empathy & Before/after video & 12.80 & 0.0004 *** & 0.0138 \\ 
(Q1-12)& Appearance $\times$ Self-disclosure & 0.0713 & 0.9312 \em{ns} & 0.0002 \\ 
& Appearance $\times$ Before/after video & 0.2378 & 0.6259 \em{ns} & 0.0003 \\ 
& Self-disclosure $\times$ Before/after video & 30.20 & 0.0000 *** & 0.0621 \\
& Appearance $\times$ Self-disclosure $\times$ Before/after video & 0.3279 & 0.7205 \em{ns} & 0.0007 \\ 
\hline
& Appearance & 0.5355 & 0.4645 \em{ns} & 0.0006 \\ 
& Self-disclosure & 8.248 & 0.0003 *** & 0.0178\\ 
Affective & Before/after video & 43.03 & 0.0000 *** & 0.0451 \\ 
empathy & Appearance $\times$ Self-disclosure & 0.6513 & 0.5216 \em{ns} & 0.0014 \\ 
(Q1-6) & Appearance $\times$ Before/after video & 0.0655 & 0.7981 \em{ns} & 0.0001 \\ 
& Self-disclosure $\times$ Before/after video & 15.73 & 0.0000 *** & 0.0333 \\ 
& Appearance $\times$ Self-disclosure $\times$ Before/after video & 1.107 & 0.3309 \em{ns} & 0.0024 \\ 
\hline
& Appearance & 2.234 & 0.1354 \em{ns} & 0.0024 \\ 
& Self-disclosure & 6.421 & 0.0017 ** & 0.0139\\ 
Cognitive & Before/after video & 0.6771 & 0.4108 \em{ns} & 0.0007 \\ 
empathy & Appearance $\times$ Self-disclosure & 0.6893 & 0.5022 \em{ns} & 0.0015 \\ 
(Q7-12) & Appearance $\times$ Before/after video & 1.220 & 0.2697 \em{ns} & 0.0013 \\ 
& Self-disclosure $\times$ Before/after video & 27.27 & 0.0000 *** & 0.0564 \\ 
& Appearance $\times$ Self-disclosure $\times$ Before/after video & 0.2472 & 0.7810 \em{ns} & 0.0005 \\ 
\hline
Empathic & Appearance & 3.881 & 0.0491 * & 0.0042 \\ 
response & Self-disclosure & 4.312 & 0.0137 * & 0.0094\\ 
(BeQ) & Appearance $\times$ Self-disclosure & 2.314 & 0.0995 \em{ns} & 0.0050 \\ 
\hline
\end{tabular}} 

\em{p}:
{{*}p\textless\em{0.05}}
{{**}p\textless\em{0.01}}
{{***}p\textless\em{0.001}}
\label{table2}
\end{adjustwidth}
\end{table}
\renewcommand{\arraystretch}{1.0}

\begin{figure}[tbp]
    \hspace{-4.75cm}
    \includegraphics[scale=0.3]{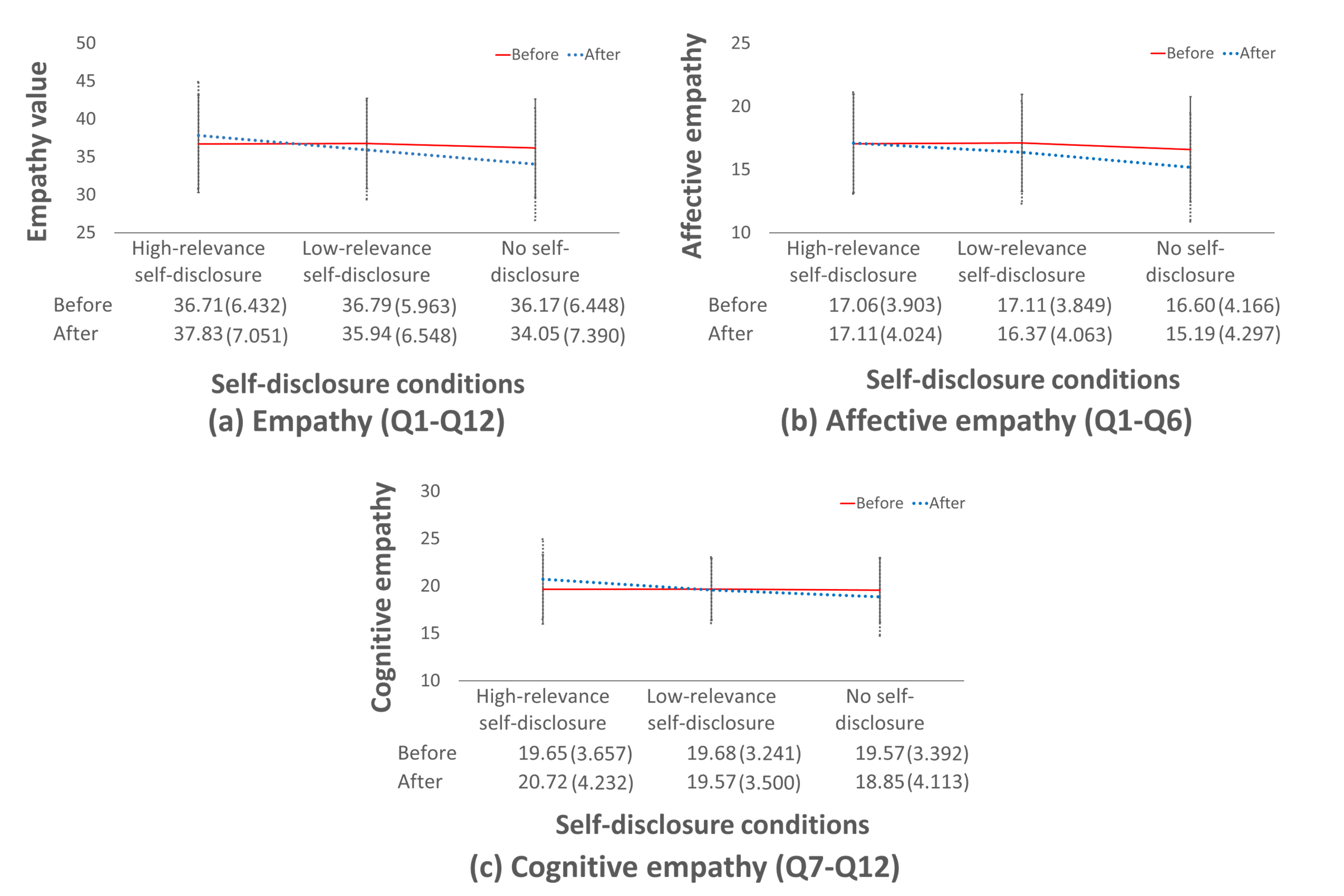}
    \caption{All graphs of the interaction between self-disclosure and before/after video}
    \label{fig5}
\end{figure}

\renewcommand{\arraystretch}{1.1}
\begin{table}[tbp]
\caption{Analysis results of multiple comparison(simple main effect)}
\vspace{1mm}
\begin{adjustwidth}{-1.0in}{0in} 
\scalebox{1.0}{
\begin{tabular}{cllll}\hline
& \multicolumn{1}{c}{Factor} & \multicolumn{1}{c}{\em{F}} & \multicolumn{1}{c}{\em{p}} & \multicolumn{1}{c}{$\eta^2_p$}\\ \hline
& Self-disclosure at Before video & 0.8858 & 0.4127 \em{ns} & 0.0019 \\ 
& Self-disclosure at After video & 22.34 & 0.0000 *** & 0.0467\\ 
Empathy & Before/after video at High-relevance self-disclosure & 16.54 & 0.0001 *** & 0.0516 \\
(Q1-12) & Before/after video at Low-relevance self-disclosure & 10.01 & 0.0017 ** & 0.0319 \\ 
& Before/after video at No self-disclosure & 38.52 & 0.0000 *** & 0.1125 \\ 
\hline
& Self-disclosure at Before video & 1.512 & 0.2211 \em{ns} & 0.0033 \\ 
Affective & Self-disclosure at After video & 16.75 & 0.0000 *** & 0.0354 \\ 
empathy & Before/after video at High-relevance self-disclosure & 0.0927 & 0.7610 \em{ns} & 0.0003 \\ 
(Q1-6) & Before/after video at Low-relevance self-disclosure & 18.43 & 0.0000 *** & 0.0571 \\ 
& Before/after video at No self-disclosure & 46.25 & 0.0000 *** & 0.1321 \\ 
\hline
& Self-disclosure at Before video & 0.0943 & 0.9100 \em{ns} & 0.0002 \\ 
Cognitive & Self-disclosure at After video & 17.36 & 0.0000 *** & 0.0367 \\ 
empathy & Before/after video at High-relevance self-disclosure & 41.35 & 0.0000 *** & 0.1197 \\
(Q7-12) & Before/after video at Low-relevance self-disclosure & 0.5183 & 0.4721 \em{ns} & 0.0017 \\ 
& Before/after video at No self-disclosure & 13.39 & 0.0003 ** & 0.0422 \\ 
\hline
\end{tabular}} 

\em{p}:
{{*}p\textless\em{0.05}}
{{**}p\textless\em{0.01}}
{{***}p\textless\em{0.001}}
\label{table3}
\end{adjustwidth}
\end{table}
\renewcommand{\arraystretch}{1.0}

Table~\ref{table2} shows the results of an ANOVA for the 12-item questionnaire. 
It also shows the results of an ANOVA for affective empathy (Q1-Q6) and cognitive empathy (Q7-Q12), which are classifications of empathy. 
The results are summarized in this paper, focusing on the areas with simple main effects where the interaction was significant.
Also, we investigated the results of an analysis done to judge the empathic response, which was surveyed only after the video was watched. 
For multiple comparisons, we examined the existence of significant differences by using Holm's multiple comparison test. 
\\ \indent
From the results of each questionnaire, a significant difference was found in the interaction between two factors, self-disclosure and before/after video. 
The results of the interaction are shown in Figure~\ref{fig5}.
This graph also shows the mean and S.D. for each condition.
Also, there was no significant interaction between the appearance factor and the self-disclosure factor under all conditions. 
Below, items for which an interaction effect was found are not discussed even if a main effect was found. 
For items for which no interaction was found and a main effect was observed, the result of the main effect is shown. 
Therefore, we investigated the simple main effects for the factors of self-disclosure and before/after video watching. 
Table~\ref{table3} shows the results of multiple comparison for the 12-item questionnaire. 

\subsection*{Empathy value}
\begin{figure}[tbp]
    \includegraphics[scale=0.3]{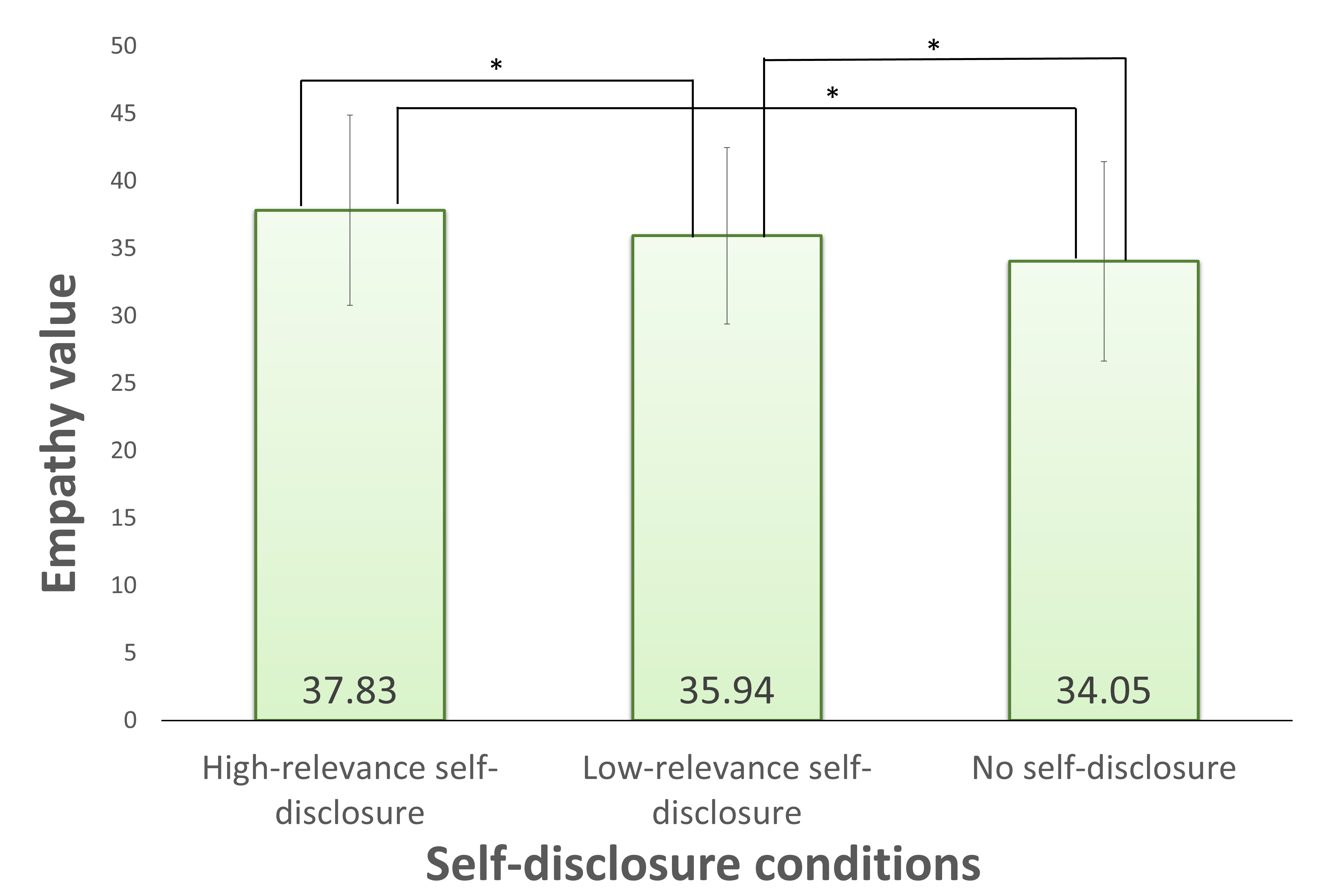}
    \caption{Results of multiple comparison for self-disclosure after watching video for empathy. Error bars show standard deviation.}
    \label{fig6}
\end{figure}

The results for empathy (Q1-12) showed an interaction between the factors of self-disclosure factor and before/after video watching. 
The main effects of the self-disclosure factor and the before/after video factor were also significant, but they were omitted because of the interaction effect that self-disclosure and watching the video factor. 
\\ \indent
As a result of a multiple comparison, the simple main effect of the self-disclosure factor after watching the video showed a significant difference in the combination of all three levels as shown in Figure.~\ref{fig6}. 
In addition, the simple main effect before/after video for each self-disclosure condition was significantly different before/after video under all self-disclosure conditions. 
On the basis of the above results, it was suggested that self-disclosure facilitated empathy when the relevance was high and that empathy was suppressed when there was no self-disclosure. 
In addition, in the case of less relevant self-disclosure, empathy was suppressed. High-relevance self-disclosure was most likely to facilitate empathy, and no self-disclosure suppressed it. 
From the results of the post-hoc analysis, it was found that self-disclosure was effective for empathy.

\subsection*{Affective empathy}
\begin{figure}[tbp]
    \includegraphics[scale=0.3]{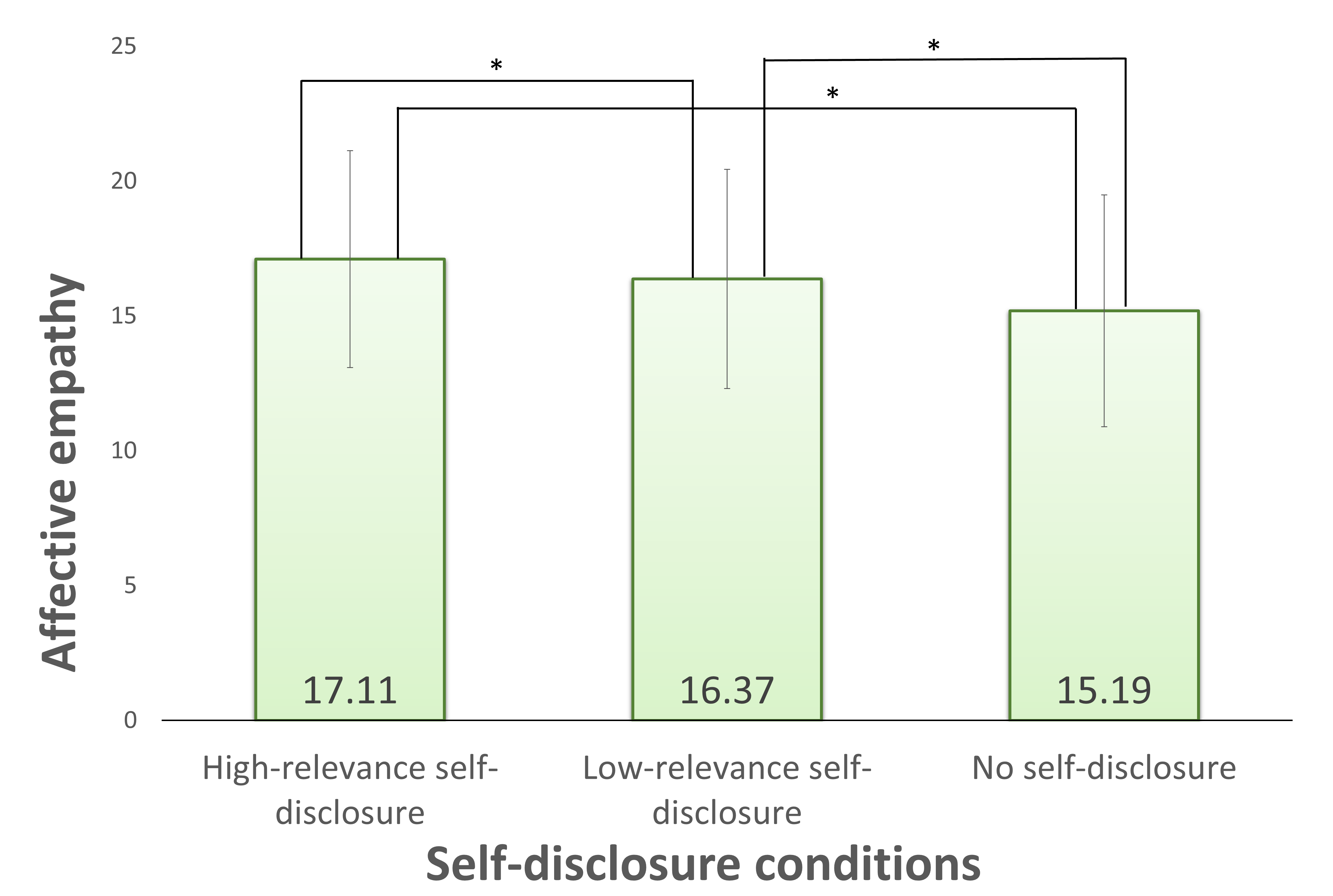}
    \caption{Results of multiple comparison for self-disclosure after watching video for affective empathy. Error bars show standard deviation.}
    \label{fig7}
\end{figure}

Similarly, the results for affective empathy (Q1-Q6) showed an interaction between the self-disclosure factor and before/after video. 
The main effects of the self-disclosure factor and the before/after video factor were also significant, but they were omitted because of the interaction effect that self-disclosure and watching the video factor. 
\\ \indent
As a result of a multiple comparison, the simple main effect of the self-disclosure factor after watching the video showed a significant difference in the combination of all three levels as shown in Figure~\ref{fig7}.
However, the simple main effect after watching the video for each self-disclosure condition was not significantly different between before/after video with high relevance. 
Under the other self-disclosure conditions, a significant difference was observed before/after video, and the result was that affective empathy was suppressed. 
This suggests that affective empathy is not suppressed only in the case of high-relevance self-disclosure. 
From the results of the post-hoc analyses, it was found that self-disclosure was effective for affective empathy.

\subsection*{Cognitive empathy}
\begin{figure}[tbp]
    \includegraphics[scale=0.3]{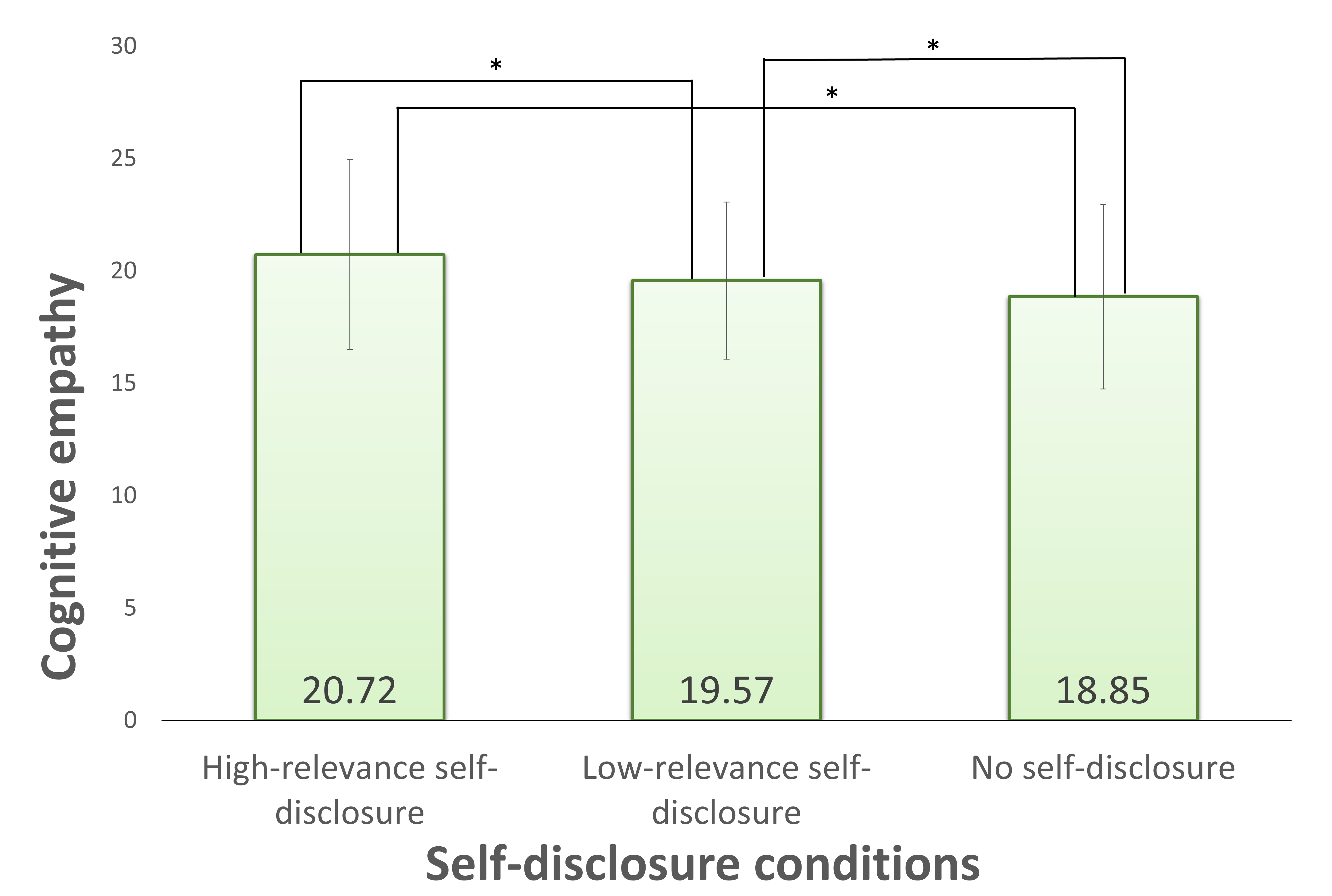}
    \caption{Results of multiple comparison for self-disclosure after watching video for cognitive empathy. Error bars show standard deviation.}
    \label{fig8}
\end{figure}

In addition, the results for cognitive empathy (Q7-Q12) showed an interaction between the self-disclosure factor and before/after video. 
The main effect of the self-disclosure factor was also significant but was omitted because of the interaction effect that self-disclosure and watching the video factor. 
\\ \indent
As a result of a multiple comparison, the simple main effect of the self-disclosure factor after watching the video showed a significant difference in the combination of all three levels as shown in Figure~\ref{fig8}. 
However, the simple main effect after watching the video for each self-disclosure condition was not significantly different between before/after video with low-relevance. 
Under the other self-disclosure conditions, self-disclosure facilitated cognitive empathy when the relevance was high, and no self-disclosure suppressed cognitive empathy. 
From the above, it was suggested that high-relevance self-disclosure facilitated cognitive empathy and that no self-disclosure suppressed cognitive empathy. 
From the results of the post-hoc analyses, it was found that self-disclosure was effective for cognitive empathy.

\subsection*{Empathic response}
\begin{figure}[tbp]
    \includegraphics[scale=0.3]{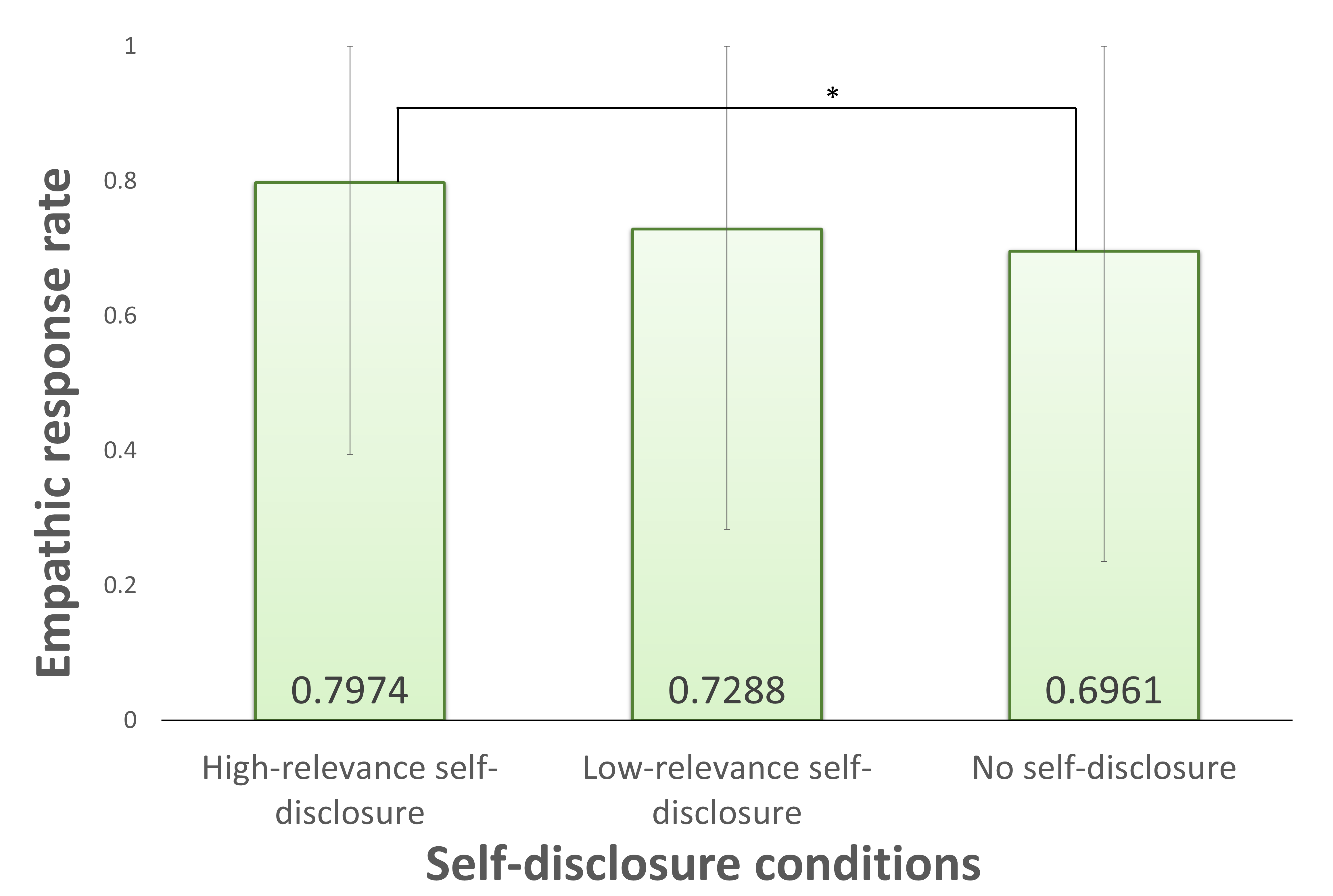}
    \caption{Results for self-disclosure for empathic response. Error bars show standard deviation.}
    \label{fig9}
\end{figure}

Finally, the results for empathic response showed no interaction between the appearance and self-disclosure factors. 
The main effects of the appearance factor and the self-disclosure factor were also significant.
\\ \indent
As a result, the main effect of the appearance factor showed a significant difference in the two levels (human-like: mean = 0.7691, S.D. = 0.4219, robot mean = 0.7124, S.D. = 0.4531). 
The main effect of the self-disclosure factor showed a significant difference between high-relevance and no self-disclosure as shown in Figure~\ref{fig9} (high-relevance: mean = 0.7974, S.D. = 0.4026, low-relevance: mean = 0.7288, S.D. = 0.4453, no self-disclosure: mean = 0.6961, S.D. = 0.4607). 

\section*{Discussion}
\subsection*{Supporting Hypotheses}
The way to improve the relationship between humans and anthropomorphic agents is to have humans empathize with the agents.
This idea is supported by several previous studies\cite{Gaesser13,Klimecki16}.
Human empathy for agents is a necessary component for agents to be used in society.
When agents are able to take an appropriate approach to human empathy, humans and agents can build a trusting relationship.
\\ \indent
In this study, the experiments were conducted to investigate the conditions necessary for humans to empathize with anthropomorphic agents. 
We focused on agent appearance and agent self-disclosure as factors that influence human empathy.
The purpose of this study is to investigate whether humans can elicit more empathy when they make a self-disclosure related to a particular situation in an interaction with an empathy agent. 
For this purpose, we formulated two hypotheses and analyzed the data obtained from the experiments. 
\\ \indent
The results supported H1 in that, among the three types of self-disclosure (high-relevance self-disclosure, low-relevance self-disclosure, no self-disclosure) from the empathy agent, high-relevance self-disclosure is most likely to facilitate empathy, and no self-disclosure suppresses empathy.
We hypothesized that empathy was facilitated only by high-relevance self-disclosure and that empathy suppressed no self-disclosure. However, low-relevance self-disclosure suppressed empathy. 
\\ \indent
Next, the experiments supported H2 in that in interacting with agents, appearance factors have little impact on promoting empathy through self-disclosure. 
So far, appearance and self-disclosure have been studied for human empathy. 
There is a reason for the choice of the appearances we used this time. 
For both of the agents, we adopted a body structure similar to that of humans on the premise that the agents were doing the same work as humans. 
It should be noted that there was no interaction between appearance and self-disclosure for an experiment in which different models were prepared for the appearance but the conditions for self-disclosure were set. 

\subsection*{Comparison with previous studies}
Shaffer et al.\cite{Shaffer19} asked participants to imagine a pregnant woman smoking and to write down the reasons why she smokes.
The results showed that participants empathized with the pregnant woman more after writing than before.
In our study, instead of writing, the agents self-disclosed.
In addition, instead of the pregnant woman, we investigated the impact of human empathy on the agent's appearance (human-like appearance or robot appearance).
The results showed a similar trend to previous studies, as human empathy was affected by the agent's self-disclosure regardless of appearance.
\\ \indent
Pan et al.\cite{Pan20} investigates human impressions of machines by the level of self-disclosure of the machines.
The self-disclosure efforts elicited information and politeness from the humans.
Our study also showed a similar trend in that the promotion of empathy was observed.
Similarly, Lee et al.\cite{Lee20} improved participants' self-disclosure, intimacy, and enjoyment when the chatbot self-disclosed.
In this study, agent self-disclosure promoted human empathy when it was relevant to the scenario.
The results showed that the impact on empathy changed by the relevance of self-disclosure.
\\ \indent
Riek et al.\cite{Riek09} investigated the appearance that a robot needs when interacting with a human.
The results showed that the robot was more likely to be empathetic if it had a human-like shape.
In this study, human empathy was similarly effective for human-like appearance and robot appearance because the robots had an appearance similar to human structure.
However, since this experiment used two different human-like appearances, it is necessary to consider the influence of anthropomorphism.
Anthropomorphism can affect interaction with humans and can affect trust\cite{Damiano18,Geraci21,Geraci21-2}.
\\ \indent
In our study, it was found that the self-disclosure factor promotes human empathy toward an anthropomorphic agent. 
In addition, an interaction was not observed between the appearance factor and the self-disclosure factor. 
We believe that this study will be an important one that separates appearance and self-disclosure as separate factors.
\\ \indent
Although this study focused on promoting empathy, it was confirmed from the results of that empathy was suppressed. 
This result has not been found in other studies.
By properly using empathy depending on the situation, we think that the impact on humans can be adjusted for anthropomorphic agents introduced into human society in the future.

\subsection*{Empathic response}
We discuss the results for behavior as an analysis of empathic response. 
In the experiments, participants played the role of observer for the target empathy agent. 
Observers responded empathetically to any information available from the target. 
The choice of whether to lend money was considered to be empathic response behavior. 
As a result of analyzing the behavior related to the empathic response after watching the video, a significant difference was found. 
However, unlike the other analyses, the effect size was small for empathic responses, and thus, the effect on empathic responses was small in the experiments. 
We think that this did not affect the behavior because the interaction time between the empathy agent and the participants was as short as three minutes.

\subsection*{Limitations}
As a limitation of this study, participants interacted with the empathy agents by watching a video. 
The current results are not enough because the sense of distance is different from the case of agents actually introduced into society. 
We will proceed with research in an environment where participants and anthropomorphic agents actually interact with each other. 
\\ \indent
We need to discuss low-relevance self-disclosure. 
It was observed from a manipulation check that the content of the self-disclosure itself is self-disclosure. 
However, empathy is not always suppressed when relevance is low. 
In the experiments, empathy was suppressed, but it may be facilitated depending on the content, even with low-relevance self-disclosure. 
Therefore, it is possible that empathy may be facilitated by the type and content of self-disclosure, which is different from the low-relevance self-disclosure we used. 
For this reason, it is effective to have self-disclosure that is high in relevance when humans empathize with empathy agents and to not allow self-disclosure if humans do not want to empathize. 
When introducing empathy agents using self-disclosure that is low in relevance, it should not be important whether humans have empathy for the agents.
\\ \indent
In addition, in this study, the appearance factors were roughly divided into two types. 
However, if a suitable appearance is prepared for each situation, it is possible that an interaction between the appearance factor and the self-disclosure factor may be observed. 
However, appearance factors vary greatly depending on human taste, and humans themselves do not have exactly the same appearance. 
Therefore, anthropomorphic agents should not be judged by their fixed appearance.

\section*{Conclusion}
To solve the problem of agents not being accepted by humans, we hope that agents will be used more in human society in the future by having humans empathize with them. 
This study is an example of how empathy can be facilitated between humans and agents. 
The experiment was conducted with a three-factor mixed-plan, and the number of between-participant factors was two, appearance and self-disclosure, and the within-participant factor was the empathy values before/after video to measure the change in empathy. 
The number of levels of each factor was two for appearance (human, robot), three for self-disclosure (high-relevance self-disclosure, low-relevance self-disclosure, no self-disclosure), and two for stimulation (before, after). 
The dependent variable was the empathy that the participants had. 
As a result, we found that the appearance factor did not have a main effect, and self-disclosure, which is highly relevant to the scenario used, facilitated more human empathy with statistically significant difference. 
We also found that no self-disclosure suppressed empathy. In addition, self-disclosure was found to be important for manipulating empathy toward the other party. 
These results support our hypotheses. Moreover, the empathic response was affected by appearance and self-disclosure factors. 
This study is an important example of how human empathy can work for artifacts. 
Agents, which are increasingly used in human society, have been found to gain empathy from humans through self-disclosure. 
As future research, we can develop empathy agents for various situations by considering cases in which we can strengthen or weaken a specific empathy element for affective empathy and cognitive empathy.

\section*{Supporting information}



\paragraph*{S1 File.}
\label{S1_File}
{\bf Complete data set}  

\paragraph*{S2 File.}
\label{S2_File}
{\bf Complete scenario contents}  

\paragraph*{S1 Video.}
\label{S1_Video}
{\bf human-like agent ver.}  

\paragraph*{S2 Video.}
\label{S2_Video}
{\bf robot agent ver.}




\section*{Author Contributions}
{\bf Conceptualization:} Takahiro Tsumura, Seiji Yamada. \\ \noindent
{\bf Data curation:} Takahiro Tsumura.\\ \noindent
{\bf Formal analysis:} Takahiro Tsumura.\\ \noindent
{\bf Investigation:} Takahiro Tsumura.\\ \noindent
{\bf Methodology:} Takahiro Tsumura, Seiji Yamada.\\ \noindent
{\bf Project administration:} Takahiro Tsumura.\\ \noindent
{\bf Resources:} Takahiro Tsumura.\\ \noindent
{\bf Software:} Takahiro Tsumura.\\ \noindent
{\bf Supervision:} Takahiro Tsumura, Seiji Yamada.\\ \noindent
{\bf Validation:} Takahiro Tsumura.\\ \noindent
{\bf Visualization:} Takahiro Tsumura.\\ \noindent
{\bf Writing – original draft:} Takahiro Tsumura.\\ \noindent
{\bf Writing – review \& editing:} Takahiro Tsumura.

\nolinenumbers

%
%
%

\bibliography{test}

\end{document}